\documentclass[useAMS,usenatbib]{mn2e}

\usepackage[psamsfonts]{amssymb}
\usepackage[dvips]{graphicx}
\usepackage{amsmath,alltt}
\usepackage{multirow}
\usepackage{rotating}
\usepackage{lscape}

\title[Small-N Collisional Dynamics]{Small-N Collisional 
Dynamics:  Pushing Into the Realm of Not-So-Small-N}
\author[Leigh \& Geller]{Nathan
  Leigh$^{1}$, Aaron M.~Geller$^{2}$\thanks{E-mail:
    leighn@mcmaster.ca (NL); a-geller@northwestern.edu (AG)} \\
$^{1}$European Space Agency, Space Science Department, Keplerlaan 1,
2200 AG Noordwijk, The Netherlands \\
$^{2}$Center for Interdisciplinary Exploration and Research in Astrophysics (CIERA) and Dept. of Physics and Astronomy, \\ Northwestern University, 2145 Sheridan Rd, Evanston, IL 60208, USA}
\begin{document}

\pagerange{\pageref{firstpage}--\pageref{lastpage}} \pubyear{2011}

\maketitle

\label{firstpage}

\begin{abstract}
In this paper, we study small-$N$ gravitational dynamics involving 
up to six objects.  We perform a large suite 
of numerical scattering experiments involving single, 
binary, and triple stars.  This is done using the FEWBODY numerical 
scattering code, which we have upgraded to treat encounters involving 
triple stars.  We focus on outcomes that result in 
direct physical collisions between stars, 
within the low angular momentum and high absolute orbital energy 
regime.  The dependence of the 
collision probability on the number of objects involved in the 
interaction, $N$, is found for fixed total energy and angular 
momentum.  Our results are consistent with a 
collision probability that increases approximately as $N^2$.  
Interestingly, this is also what is expected from the mean 
free path approximation in the limit of very large $N$.  A more 
thorough exploration of 
parameter space will be required in future studies to 
fully explore this potentially intriguing connection.  
This study is meant as a first step in an on-going effort 
to extend our understanding of small-$N$ collisional dynamics beyond 
the three- and four-body problems and into the realm of larger-$N$.
\end{abstract}


\section{Introduction} \label{intro6}

The gravitational three-body problem was first studied by Sir Isaac
Newton in his Principia \citep{newton1686}.  After solving the
two-body problem, 
Newton boldly added a third body into the mix and attempted to create
a similar mathematical formalism to describe the motion of three
celestial bodies under their mutual gravitational attraction.  The
solution evaded Newton until his end, leaving the problem at the mercy
of his disciples.  It was later taken up by Euler \citep{euler1772}, 
Lagrange \citep{lagrange1811}, Jacobi \citep{jacobi1836},
Poincare \citep{poincare1892}, Hill \citep[e.g.][]{hill1878}, 
Henon \citep[e.g.][]{henon69}, 
and a host of others, often with a focus on the
Earth-Moon-Sun system \citep[e.g.][]{valtonen06}.  

Despite the considerable efforts of numerous researchers, a simple 
analytic solution to the general three-body problem has never been found.  
\citet{sundman1912} discovered a uniformly convergent infinite series 
involving familiar functions that technically solves the three-body 
problem.  However, in order to attain a reasonable level of accuracy, 
the solution must contain on the order of $10^{8,000,000}$ terms 
\citep{valtonen06}.  More practical solutions require a number of 
simplifying assumptions to make the general three-body problem 
tractable \citep[e.g.][]{henon69}.  As a result, the most useful 
analytic solutions tend to be solely applicable to a very narrow 
subset of the total allowed parameter space.  

The introduction of computers within the last few decades 
revolutionalized our understanding of the three-body problem.  This 
allowed for the direct integration of 
the equations of motion for each particle over small time steps, 
incrementally moving each body forward in an iterative fashion.
Despite the fact that this approach completely
transformed our tool-set for studying the three-body problem, it too has
its short-comings.  
For example, the times required to run the
simulations to completion are often quite long.  The precise definition 
of a ``complete encounter'' can often be ambiguous as well.  Quasi-stable 
configurations can form that remain bound for millions or even billions 
of years before eventually breaking apart \citep[e.g.][]{mikkola86}.  
There is also the issue of errors in computing the trajectories of the 
particles in position-space which are introduced at each time-step.  These
arise as a result of moving one body forward at a time, instead of
moving all bodies 
simultaneously.  Such errors can be minimized by taking suitably
short time-steps.  However, this comes at the often considerable cost of
simulation run-time.

When only three bodies are involved, qualitative descriptions
of the outcomes of dynamical interactions are often 
relatively straight-forward.  They include ionizations, 
exchanges, fly-bys, and collisions \citep[e.g.][]{hut83b, mikkola83, 
mikkola84b, hut84, mcmillan96, fregeau04}.  
However, the number of possible outcomes quickly increases
with increasing $N$, where $N$ is the total number of objects involved 
in the interaction \citep{leigh11}.  This complicates descriptions of the interactions, 
and introduces a considerable 
challenge in developing a physical understanding of the evolution of 
the system.  For example, nearly 100 generic outcomes are 
possible for encounters involving six objects.  
Additional bodies not only increase the parameter space 
to be explored, they also increase integration times for 
computer simulations.  Consequently, most previous numerical 
scattering studies considered only single-binary and, to a lesser 
extent, binary-binary encounters \citep[e.g.][]{heggie75, hut83a, 
mcmillan86, sigurdsson93, davies95, bacon96, fregeau04}.  

Many of the numerical scattering studies conducted to date considered 
equal-mass particles, and devoted their attention to studying the effects 
of varying the initial relative velocity or impact parameter 
\citep[e.g.][]{hut83b}.  For example, \citet{hut83a} found analytic 
approximations for high-velocity encounters, and provided 
simple formulae for the corresponding cross-sections for exchanges 
and ionizations to occur.  Similar analytic formulae were derived 
by \citet{mikkola84a} for encounters involving two binaries having 
unequal orbital energies but equal masses.  

An extensive analysis that considered unequal mass particles 
was conducted by \citet{sigurdsson93}, who 
studied the effects of dynamics on the stellar and remnant 
populations in a globular cluster.  A number of other scattering 
experiments were 
designed purely to study the formation of different types of 
stellar exotica in globular clusters, including blue stragglers 
\citep[e.g.][]{leonard89}, low-mass x-ray binaries 
\citep[e.g.][]{sigurdsson95}, cataclysmic variables 
\citep[e.g.][]{ivanova06}, and millisecond pulsars 
\citep[e.g.][]{ivanova08}.  Many of these also considered a range 
of particle masses.  The effects of general relativity have also 
been studied in the context of the three- and four-body problems.  
For example, \citet{mikkola90} and \citet{valtonen94} considered 
interactions involving binary super-massive black holes during 
the mergers of galaxies, and identified several important trends 
arising from energy losses due to gravitational radiation.

In this paper, we study gravitational interactions involving up to six 
objects.  Our goal is to better understand how the outcome of an 
encounter depends on the number of interacting objects.  
This is a first step toward extending techniques developed for the 
three-body problem, where the vast majority of research efforts have 
thus far been concentrated, to treat larger-$N$ encounters.  To this end, 
we perform $>10^7$ numerical scattering 
experiments involving single, binary and triple star systems.  The 
number of possible encounter outcomes is a steeply increasing function 
of $N$.  This presents a considerable challenge when trying to draw parallels 
between encounters involving different numbers of objects.  To minimize 
this issue, we focus 
only on outcomes resulting in direct physical collisions, which we 
consider to have occurred when the radii of any two stars overlap.  

In Section~\ref{method}, we describe the set-up for our numerical scattering 
experiments, including the range of initial conditions considered.  
Additionally, we develop an analytic formula for the collision probability as a function 
of $N$, that is based on a simple analytic model originally derived for 1+2 encounters. 
In Section~\ref{results}, we present the results of our analysis of this large suite of 
single-binary (1+2), binary-binary (2+2), single-triple (1+3), 
binary-triple (2+3), and triple-triple (3+3) scattering 
experiments.  Here, we fit the analytic formula to the results from these numerical 
scattering experiments, thereby deriving the $N$-dependence of 
the collision probability.  
The implications of our analysis for small-$N$ collisional dynamics 
are discussed in Section~\ref{discussion}.  Concluding remarks are given in 
Section~\ref{summary}.  
Finally, in an appendix, we present a formalism for creating schematic 
diagrams that quantitatively depict the 
evolution of an interaction in energy-space, and briefly discuss some 
of their possible applications. 

\section{Method} \label{method}

In this section, we describe the details of our numerical scattering 
experiments, and present the general form for the collision 
probability to which our results will be compared in Section~\ref{results}.

\subsection{Scattering Experiments} \label{scattering}

We calculate the outcomes of a series of single-binary, binary-binary, 
single-triple, binary-triple, and triple-triple 
encounters using the FEWBODY\footnote{for the source code, see 
http://fewbody.sourceforge.net} numerical scattering code.  The code 
integrates the usual $N$-body equations in configuration- (i.e. position-) 
space in order to advance the system forward in time.  For details 
concerning the adaptive integration, 
classification techniques, etc. 
used by FEWBODY, we refer the reader to \citet{fregeau04}.

We adapted the FEWBODY code to handle all encounters involving
singles, binaries, and triples.  (Specifically we created additional subroutines 
to simulate 1+3 and 3+3 encounters\footnote{The authors are happy to 
provide these additional subroutines to users of FEWBODY upon request.  They 
will be made available on the FEWBODY homepage shortly, along with the 
latest version of the source code (FEWBODY 0.27).};
codes to simulate encounters between binaries and singles only, as well as a 
2+3 encounter code, were previously available in the FEWBODY package.)  
We use the same criteria as \citet{fregeau04} to decide when a given 
encounter is complete, defined as the point at which the 
separately bound hierarchies that make up the system are no longer 
interacting with each other or evolving internally.  

To perform physical collisions between stars, FEWBODY uses the 
``sticky-star'' approximation.  This treats stars as rigid spheres 
with radii equal to their stellar radii.  A physical collision is 
assumed to occur when the radii of the stars overlap.  When this 
happens, the stars are merged assuming conservation of linear 
momentum and no mass loss.  This does not account for tidal effects, 
which could significantly increase the collision probability 
\citep[e.g.][]{mcmillan86}, but are beyond the scope of this work.  For 
this reason, we consider the collision probabilities presented in this 
paper to be lower limits for the true values.

Previous scattering experiments have shown that the total energy 
and angular momentum are the most important parameters in deciding the 
outcomes of 1+2 interactions \citep[e.g.][]{valtonen06}.  Therefore, we will 
fix the total energy and angular momentum 
when comparing between encounters involving different numbers of objects.  
By fixing these quantities, we aim to remove the dependences of the encounter outcomes
on energy and angular momentum (when comparing between different encounter 
types), thereby normalizing the comparisons to reveal the $N$-dependence 
of the collision probability.  
We consider several different combinations of the total energy 
and angular momentum, which we henceforth refer to as Runs (see below).  For 
each of these combinations or Runs, both the total energy and angular momentum 
are always chosen to be the same to within a factor of $\sim 2$ for every type 
of encounter.  

We focus on three primary Runs (labeled Runs 1, 2, and 3 in 
Table~\ref{table:initial-conditions}) in Section~\ref{results}.  
However, in order to check the robustness of our results, we also perform 
eight additional Runs with similar total energy 
and angular momentum as adopted in Runs 1, 2, and/or 3.  
We selected
the following additional combinations of semi-major axes for the two orbits used
in each of these additional Runs:  0.5 AU and 3.5 AU, 0.5 AU and 6.5 AU, 
0.5 AU and 8 AU, 0.5 AU and 10 AU, 1 AU and 8.5 AU, 1 AU and 11.5 AU, 
1 AU and 15 AU, 2 AU and 15 AU.  For all three primary Runs, we perform a total of 
800,000 numerical 
scattering experiments for each of the different encounter types, whereas this number 
is reduced to 200,000 for the other Runs.  In total, this amounts to 12 million 
simulations.  This number is chosen to be a balance between 
statistical significance and simulation run-time, which can be quite long for 
simulations involving five or more objects.  

As we design our 
experiment such that the total angular momentum is roughly the same for 
every encounter type within a given Run, we choose the sum of the 
semi-major axes of the 
two objects involved in the interaction (which we will call the 
\textit{geometric cross-section}) 
to be equivalent to within a factor of $\sim 2$ for every encounter 
type. These 
cross-sections are determined by the initial semi-major axes of any binaries 
and/or triples involved in the interaction, since the physical radii of the stars are small 
in comparison.  
(For triples, the cross section is primarily determined by the semi-major axis of the 
outer orbit.)  Specifically, we choose the semi-major 
axes shown in Table~\ref{table:initial-conditions}.  A semi-colon is used to 
separate different objects, whereas a comma is used to 
separate the orbits within triples.  Parantheses are used to enclose the
semi-major axes of triples, with the smaller of the two semi-major axes 
always corresponding to the inner binary.  

We assume equal masses of 1 M$_{\odot}$ and stellar radii of 1 R$_{\odot}$ 
for all stars, and circular orbits for all binaries and triples.  This is 
done to make our 
exploration of the relevant parameter space more tractable.  However we 
expect the collision probability to in general depend on the mass ratios and 
orbital eccentricties \citep[e.g.][]{sigurdsson93}.  
Unequal masses and non-zero eccentricities are 
beyond the scope of the present paper.  However we intend to address these 
issues in future work.  The ratio between the inner and outer semi-major axes 
is chosen to be $\ge$ 7 for all triples, since this roughly corresponds
to the stability boundary for the internal evolution of triples composed of
equal mass-particles with initially zero-eccentricity inner orbits and 
moderate-eccentricity outer orbits \citep{mardling01}.  The initial angle 
of inclination between the inner and outer orbits of all triples, in addition 
to their initial relative phases, are chosen at random.

For each Run, we populate a grid of scattering experiments varying only the 
relative velocity at infinity and the impact parameter.  
Specifically, we select values for the relative velocity at infinity 
from 0 to 1.1 in equally-spaced intervals of 0.004, 
in units of the critical velocity $v_{crit}$ (defined as the relative velocity that gives a 
total energy of zero for the encounter).  We select values for the impact parameter
from 0 to 7 in equally-spaced intervals of 1, in units of the geometric cross-section for 
collision (e.g. the semi-major axis of the binary for a 1+2 encounter, the 
sum of the semi-major axis of the outer orbit of the triple and that of the 
binary in a 2+3 encounter, etc.).  In this way, for a given impact parameter, 
we ensure that the range in both the total 
angular momentum and the total energy are equal for every encounter type to 
within a factor of $\sim 2$.  
Finally, at each point in this grid, we perform multiple scattering experiments 
randomizing all angles in the encounter, including
the angles at impact between the orbital planes of 
any binaries and triples involved in the encounters.

%\clearpage

\begin{table*}
\caption{Initial semi-major axes of all binaries and triples for every Runs 1, 2, and 3}
\begin{tabular}{|c|c|c|c|}
\hline
Encounter Type   &          Run 1            &           Run 2           &           Run 3           \\
                 &         (in AU)           &          (in AU)          &          (in AU)          \\
\hline
1+2              &            10.0             &            5.0            &             7.0            \\  
2+2              &         1.0; 10.0           &         0.5; 5.0          &          1.0; 7.0          \\
1+3              &        (1.0, 10.0)          &        (0.5, 5.0)         &         (1.0, 7.0)         \\
2+3              &     10.0; (1.0, 10.0)       &      5.0; (0.5, 5.0)      &      7.0; (1.0, 7.0)      \\
3+3              &  (1.0, 10.0); (1.0, 10.0)   &   (0.5, 5.0); (0.5, 5.0)  &   (1.0, 7.0); (1.0, 7.0)  \\
\hline
\end{tabular}
\label{table:initial-conditions}
\end{table*}

%\clearpage

\subsection{Collision Probability} \label{theory}

In this section, we present 
a general functional form for the collision probability.  
We begin by comparison to the 1+2 encounters studied in Section 8.6 of 
\citet{valtonen06}, and then generalize this formula to $N>3$.  
The best-fitting values for all free parameters 
in this analytic collision probability formula are then found 
in Section~\ref{results} 
through fits to the results of our numerical scattering experiments.  

\subsubsection{Collision Probability for $N = 3$} \label{collprob_Neq3}

The simple analytic model we use to guide our choice for the collision 
probability was originally found for resonant 1+2 interactions.  The 
model is described in detail in Section 8.6 of \citet{valtonen06}, to which
we refer the reader for its full derivation.  It is based on the 
assumption that the encounter will 
evolve via a series of successive ejections, eventually leading to the escape 
of one of the stars from the three-body system.  We use the term 
ejection to describe a scenario in which one object ends up at a great
distance from the others before falling back to resume the interaction.  
Some of these ejections are of considerably longer duration
than others.  Every ejection counts, however, since they all produce a
temporary binary independent of the duration of the ejection.  The 
temporary binary experiences close 
two-body encounters at the pericentre $q$ of every orbit 
\citep[e.g.][]{szebehely67, saslaw74, valtonen06}.  

For a 1+2 interaction, the probability that two stars will pass within a 
distance $q$ from each other can be approximated by:
\begin{equation}
\label{eqn:prob-coll}
P_{coll}(q) \sim 240C(L)q/a_0,
\end{equation}
where $a_0$ is the initial semi-major axis of the binary, and we require 
$q \ll a_0$.  The factor 
$C(L)$ is needed to account for the fact that the lifetime of 
the system depends on the total angular momentum $L$ \citep[e.g.][]{valtonen74,
anosova86}.  If we take $q$ to be the physical sizes of the objects 
involved in the encounter, 
then Equation~\ref{eqn:prob-coll} gives the collision probability for a
resonant 1+2 interaction.  It has been shown to give good agreement with 
the results of numerical scattering experiments \citep{valtonen06}.

\subsubsection{Collision Probability for $N > 3$} \label{collprob_Ngt3}

We write the collision probability for an encounter involving $N > 3$ 
objects as: 
\begin{equation}
\label{eqn:prob-coll-N}
P_{coll}(q) = {\alpha}N^{\beta}C(N,L)q/a_0,
\end{equation}
where
\begin{equation}
\label{eqn:cfunc}
C(N,L) = exp(\frac{{\delta}L}{N}) + \gamma,
\end{equation}
and $\alpha$, $\beta$, $\delta$, and $\gamma$ are all constants.  We take 
$a_0$ to be the semi-major axis of the shortest-period orbit involved in 
the interaction, and $C(N,L)$ is a 
function of both the total angular momentum $L$ and the number of objects $N$.  
As we will show 
in Section~\ref{results}, Equation~\ref{eqn:prob-coll-N} gives excellent 
agreement to the results of our numerical scattering experiments.
Below, we justify our choice for this functional 
form for the collision probability.

First, we are interested in 
quantifying the dependence of the collision probability 
on the number of objects $N$ involved in the interaction.  This is 
accounted for in Equation~\ref{eqn:prob-coll-N} by a general power-law 
dependence on $N$ (i.e. $\beta$).  Second, based on 
the results of previous numerical scattering experiments, we 
expect that an encounter outcome will depend both on the total energy 
and the total angular momentum.  Therefore, we expect that these two 
quantities should appear somewhere in the equation for the collision 
probability.  These dependences are included in 
Equation~\ref{eqn:prob-coll-N} via the terms $C(N,L)$ and $q/a_0$.  
The first term $C(N,L)$ is a function of the total angular momentum $L$, 
whereas the 
second term is roughly proportional to the total energy of the 
encounter.  This last point follows from the fact that we have defined 
$a_0$ to be the semi-major axis of the shortest-period orbit, and it is 
this orbit that has the largest absolute orbital energy.  Moreover, 
previous numerical scattering experiments of 1+2 and 2+2 encounters have 
shown that the components of the shortest-period binary involved in the 
interaction are the most likely to experience a collision during an 
encounter \citep[e.g.][]{fregeau04}.

We allow for a possible $N$-dependence in the function $C(N,L)$, since 
previous numerical scattering experiments performed 
to constrain this function considered only 1+2 interactions.  
Therefore, we do not know whether the total number of stars 
involved in the interaction will affect its lifetime, and 
play a role in determining the function $C(N,L)$.  As we will show 
in Section~\ref{results}, the specific functional form we have 
adopted for $C(N,L)$ in Equation~\ref{eqn:cfunc} is needed to ensure 
that the correct agreement with the results of our numerical scattering 
experiments persists as we move to larger total angular momentum.

It is important to note that Equation~\ref{eqn:prob-coll-N} does
not apply to 1+2 encounters.  Therefore, we do not include them 
when finding the best-fit parameters.  This is because the conditions 
imposed by our assumptions (e.g. equal masses for all stars, only 
circular orbits, etc.) are such that 1+2 interactions cannot always 
be made to fit into our normalization for comparing between the 
different encounter types without significantly modifying the initial 
parameters of the encounter.  Specifically, 1+2 encounters 
initially involve only a single bound orbit (via the binary).  However, 
all other encounter types initially involve multiple orbits, which 
affords us additional free parameters.  These can be adjusted 
to get the total energy and angular momentum to within 
our required factor of 2, and therefore define a suitable 
normalization between encounter types.  This is not always possible 
within the confines of our assumptions when 1+2 encounters are also 
included.  We will come back to this issue in Section~\ref{discussion}.

\section{Results} \label{results}

The collision probability is calculated from the output of our numerical
scattering experiments for a given encounter type and a given Run as:
\begin{equation}
\label{eqn:prob_coll_scat}
P_{coll} = \frac{N_{coll}}{N},
\end{equation}
where $N_{coll}$ is the total number of encounters that result in a direct
physical collision, and $N$ is the total number of encounters performed.  
The uncertainty for the collision
probability is calculated using Poisson statistics according to:
\begin{equation}
\label{eqn:prob_coll_err}
{\Delta}P_{coll} = \frac{\sqrt{N_{coll}}}{N}.
\end{equation}
Technically, this uncertainty should include scattering experiments that result in
unresolved outcomes \citep{hut83a}.  However, we find that the number of unresolved
outcomes is sufficiently small that it does not significantly contribute to
${\Delta}P_{\rm coll}$, and we do not include it in its calculation.  

\begin{figure}
\begin{center}
\includegraphics[width=\columnwidth]{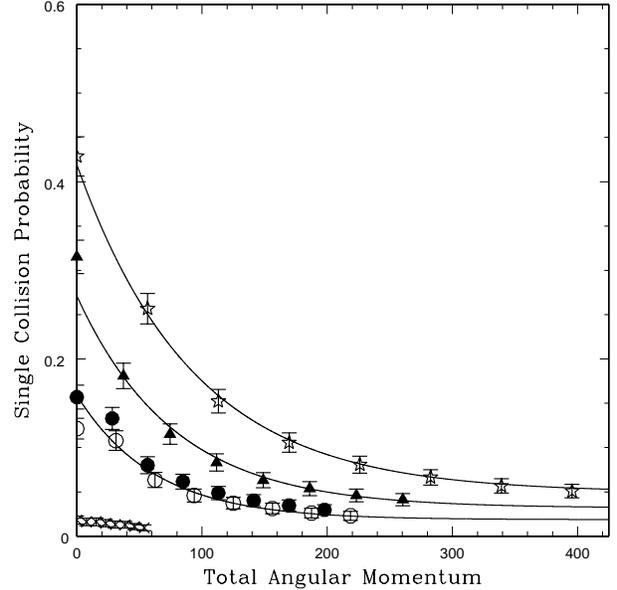}
\end{center}
\caption[The probability of a single collision occurring as a function of the total 
angular momentum for each type of encounter for Run 1]{The probability of a single 
collision occurring as a function of the total angular momentum is 
shown for Run 1 (see Table~\ref{table:initial-conditions}) 
for every type of encounter.
The open stars correspond to 3+3 encounters, the solid triangles to
2+3 encounters, the open circles to 1+3 encounters, the solid circles to
2+2 encounters, and the crosses to 1+2 encounters.  The solid lines show 
our best-fits to the data using Equation~\ref{eqn:prob-coll-N}.  
The top line corresponds to the case $N=6$, the line below
to $N=5$, and the bottom line to $N=4$.  
We show the scaled uncertainties here (and in Figures~\ref{fig:prob-merge-run2}
and~\ref{fig:prob-merge-run3}), as discussed in Section~\ref{results}. 
\label{fig:prob-merge-run1}}
\end{figure}

\begin{figure}
\begin{center}
\includegraphics[width=\columnwidth]{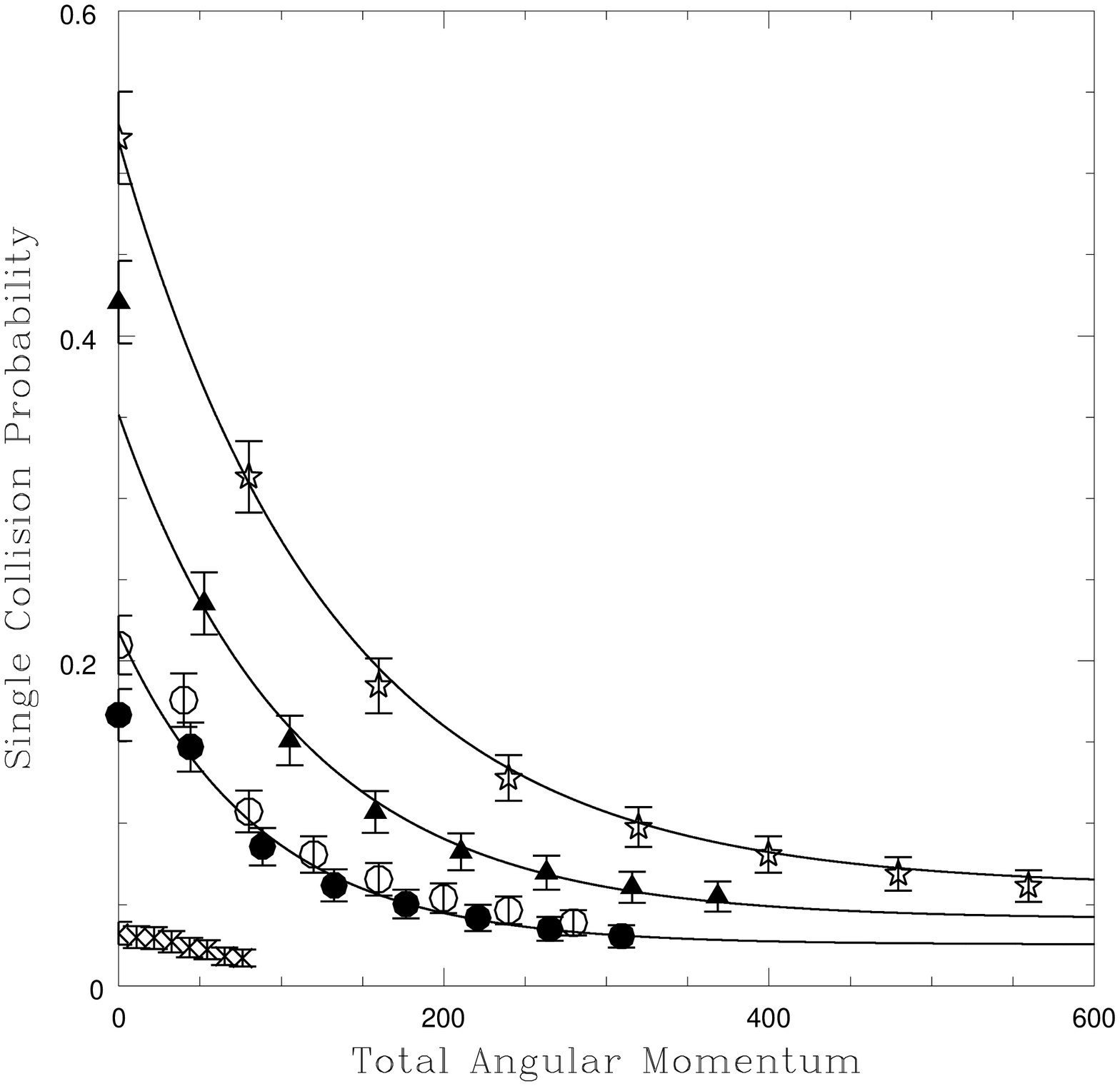}
\end{center}
\caption[The probability of a single collision occurring as a function of the total 
angular momentum for each type of encounter for Run 2]{The probability of a 
single collision occurring as a function of the total angular momentum is 
shown for Run 2 (see Table~\ref{table:initial-conditions}) 
for every type of encounter.  
The symbols representing each encounter type, as well as the lines showing 
our best-fits to the data using Equation~\ref{eqn:prob-coll-N} for each 
encounter type, are the same as in Figure~\ref{fig:prob-merge-run1}. 
\label{fig:prob-merge-run2}}
\end{figure}

\begin{figure}
\begin{center}
\includegraphics[width=\columnwidth]{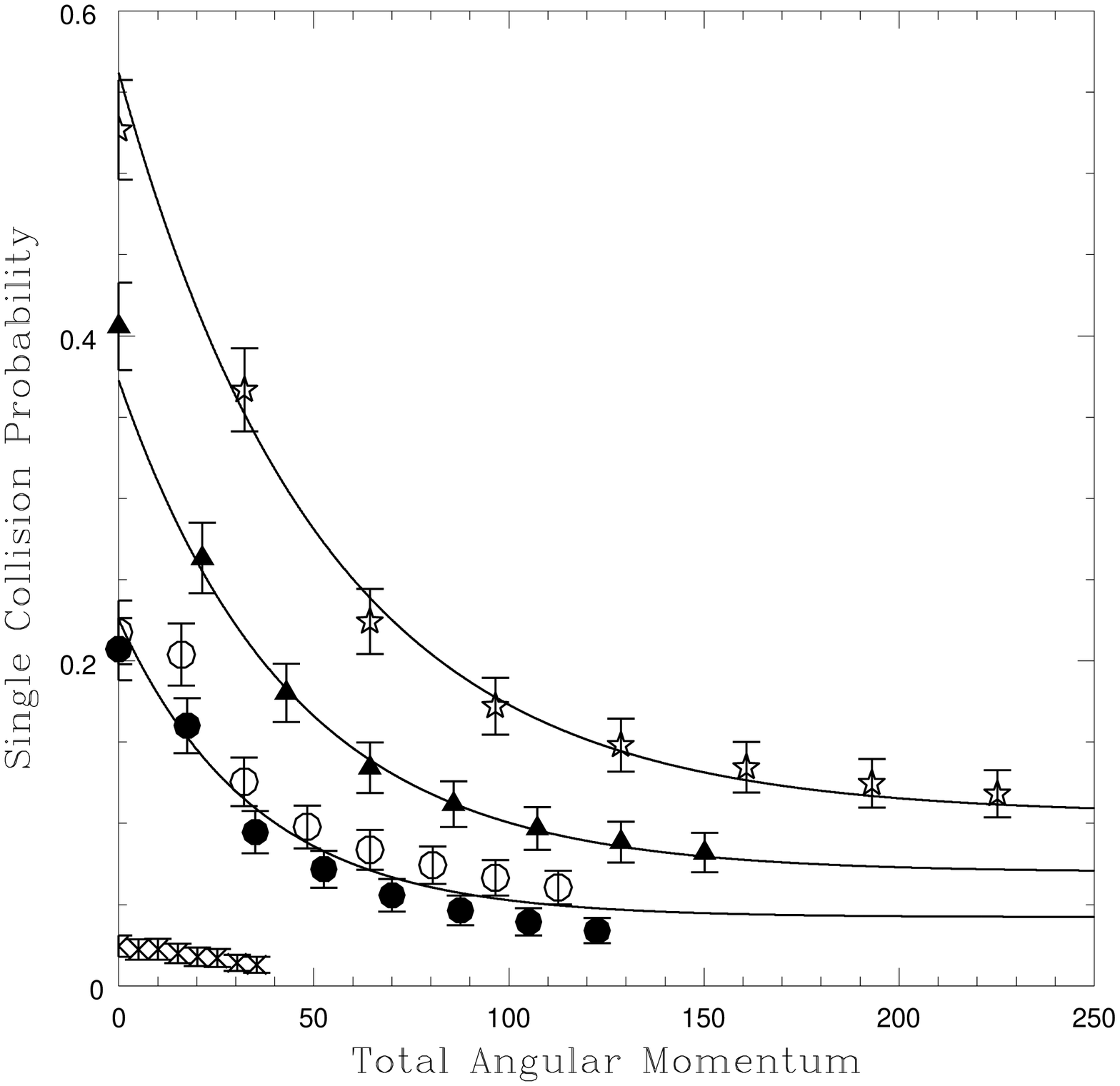}
\end{center}
\caption[The probability of a single collision occurring as a function of the total 
angular momentum for each type of encounter for Run 3]{The probability of a 
single collision occurring as a function of the total angular momentum is 
shown for Run 3 (see Table~\ref{table:initial-conditions}) 
for every type of encounter.  
The symbols representing each encounter type, as well as the lines showing
our best-fits to the data using Equation~\ref{eqn:prob-coll-N} for each 
encounter type, are the same as in Figure~\ref{fig:prob-merge-run1}.
\label{fig:prob-merge-run3}}
\end{figure}

We show these collision probabilities as a function of 
the total angular momentum for 1+2, 2+2, 1+3, 2+3, and 3+3 encounters in 
Figures~\ref{fig:prob-merge-run1}, \ref{fig:prob-merge-run2} and 
\ref{fig:prob-merge-run3} for Runs 1, 2, and 3, respectively.  The 
total angular momentum, shown on the x-axis, is provided in units of 
M$_{\odot}$$\sigma_{geo}$v$_{crit}$, where $\sigma_{geo}$ is the 
geometrical cross-section for collision in AU (which is determined by the 
semi-major axes of the orbits going into the interaction).  

We then fit Equation~\ref{eqn:prob-coll-N} to each of our individual 
Runs in order to derive the best-fitting values for the free parameters
$\alpha$, $\beta$, $\delta$, and $\gamma$,
that correctly predict the collision probability simultaneously for the
2+2, 1+3, 2+3, and 3+3 encounters.  
We use the Levenberg-Marquardt least-squares fitting technique implemented in the
MPFIT code \citep{markwardt08,more78} 
to derive these best-fitting values.  
In order to obtain realistic uncertainties on these fit parameters, we 
scale the uncertainties on our measurements by a constant factor to make the reduced 
$\chi^2$ value equal to unity.

The solid lines in 
Figures~\ref{fig:prob-merge-run1}, \ref{fig:prob-merge-run2} and 
\ref{fig:prob-merge-run3} show our fits to Equation~\ref{eqn:prob-coll-N}.  
The best-fitting parameters are shown in Table~\ref{table:bestfits} 
for all three Runs, along with their corresponding uncertainties.  

The mean power-law index for all 11 Runs is $\bar{\beta} = 2.40 \pm 0.12$.  We see weak 
evidence for an increase in our $\beta$ values with increasing geometric cross-section, 
with a Pearson's linear correlation coefficient of 0.73.  We suspect that this weak 
trend is due to our normalization beginning to break down at large impact parameters, 
which can amplify small differences in the geometric cross-sections for different 
encounters types.  The best-fitting power-law index on $N$ is consistent with a value 
of $\beta = 2$ to within roughly three standard deviations for all Runs.  Moreover, a 
chi-squared test between all 11 of our $\beta$ values and a constant power-law index 
of $\beta = 2$ returns no significant distinction, with a reduced chi-squared value 
of 1.73.  Therefore, we conclude that our results are consistent with a collision 
probability that increases approximately as $N^2$.

Additionally, we find an exponential drop-off in the collision probability 
with increasing angular momentum that is steeper for larger $N$.  Once 
again, this is the case for all Runs.  We will return to these two 
features of the collision probability relation in 
Section~\ref{discussion}. 

We find some small disagreement between the collision probabilities
for 2+2 and 1+3 encounters for Run 3.  This can be understood as 
follows.  For 1+3 encounters, the geometric cross-section for 
collision is equal to the semi-major axis of the outer orbit of 
the triple (since the radius of the single star is negligible in 
comparison).  For 2+2 encounters, however, the geometric 
cross-section is equal to the sum of the semi-major axes of the 
two binaries.  Therefore, within the framework of our experimental 
set-up, the approximation that the geometric cross-sections for 1+3 
and 2+2 encounters will be equal only holds provided one of the 
binaries in the 2+2 case has a much greater orbital separation than 
the other binary.  This assumption is a good approximation for Runs 
1 and 2, however it begins to break down for Run 3 with 
increasing impact parameter (i.e. increasing angular momentum in 
Figure~\ref{fig:prob-merge-run3}).  

\begin{table*}
\caption{Best-Fit Parameters for Equation~\ref{eqn:prob-coll-N}}
\begin{tabular}{|c|c|c|c|}
\hline
Parameter        &               Run 1               &               Run 2              &               Run 3                 \\
\hline
$\alpha$         &             1.10 $\pm$ 0.20            &             1.06 $\pm$ 0.19           &               1.73 $\pm$ 0.32               \\
$\beta$          &             2.39 $\pm$ 0.11            &             2.15 $\pm$ 0.11           &               2.25 $\pm$ 0.11               \\
$\delta$         &         -6.47e-2 $\pm$ 4.40e-3         &         -4.60e-2 $\pm$ 3.21e-3        &           -1.15e-1 $\pm$ 9.97e-3            \\
$\gamma$         &          1.33e-1 $\pm$ 1.56e-2         &          1.33e-1 $\pm$ 1.56e-2        &            2.31e-1 $\pm$ 2.33e-2            \\
\hline
\end{tabular}
\label{table:bestfits}
\end{table*}

\section{Discussion} \label{discussion}

Within the angular momentum and energy regime studied here, 
we find that the probability of a direct physical collision occurring 
during an interaction increases roughly as $N^2$.  One interpretation 
of this result can be understood as follows.  Previous numerical 
scattering experiments of 
1+2 interactions have shown that the collision probability is 
proportional to the average number of close approaches experienced by the 
system per crossing time, multiplied by the number of crossing times survived 
through \citep{valtonen06}.  
Therefore, this $N^2$ dependence may arise physically from an $N^2$ dependence in 
the total number of close approaches experienced by the system.  This can occur in 
one of (at least) two ways.  Either the number of close approaches per crossing time scales 
as $N^2$ while the number of crossing times survived through is constant, or the number of 
crossing times the system survives through (until the time when the first 
collision occurs) scales as $N^2$ while the number of close approaches per crossing 
time is constant.

We find that the average number of crossing times at the time of the first collision 
is the same to within roughly a factor of $\approx$ 2 for all encounter types, and 
that there is no trend with $N$.  
Therefore, it is unlikely that the latter scenario described above is responsible 
for producing the $N^2$ dependence.  
This suggests that perhaps the $N^2$ dependence in the collision probability may 
arise from a similar dependence in the number of close approaches per crossing time.

However, in practice, this hypothesis is much more difficult to test quantitatively.
In particular it is not immediately clear how to define a ``close approach''.  
We attempt to quantify this effect by examining animations of a handful of 
encounters in position-space, and counting the number of close 
approaches per crossing time by eye.  
It is clear that 
the number of close approaches per crossing time increases with 
increasing $N$.  However, this method is far from adequate to determine
the precise $N$-dependence of this relation at any statistically 
significant level.  

In Appendix~\ref{appendix}, we present a method that 
will improve upon this component of our analysis in future studies.  
Specifically, we present a prescription for generating schematic 
diagrams that depict the evolution of an interaction in energy-space.  
As is explicitly shown in 
Appendix~\ref{appendix},
the advantage of this technique is to provide a straight-forward means of 
defining the criterion of ``close approach'' in terms of the fraction 
of the total energy exchanged between two individual stars.  
This directly relates the definition of close approach to the 
total energy and therefore the initial conditions of the encounter, which 
is not the case if this criterion is defined purely in terms of a distance.  

Also note that, in a sense, a collision could be interpreted as a very strict definition of 
a ``close approach''. Here multiple crossing times are required before this 
definition of close approach is satisfied.  As we find that the number of crossing times until the time 
of first collision is roughly equivalent (to within a factor of $\approx$ 2) for all
encounter types, our result can potentially also be expressed as the number 
of ``close approaches'', defined in this manner, per crossing time scaling as $N^2$.  
We will explore a range of other criteria to define a close approach in a future study
and investigate explicitly the dependence of the 
number of close approaches per crossing time 
on the total number of 
stars involved in the interaction.  

Intriguingly, the collision rate, as derived from the mean free path 
approximation \citep[e.g.][]{leonard89}, also scales as $N^2$.  
In the limit of very large $N$, we would expect our relation for the 
collision probability to agree with what is predicted from the 
mean free path approximation.  On the other hand, in the limit of 
small-$N$, it is not clear that the standard assumptions of the 
mean free path approximation should still hold.  
For the angular momentum and energy regime studied here, the 
collision probability appears consistent with that of the mean free path 
approximation, at least for $N = 4, 5, 6$.  Further study is required 
to determine, e.g., how different energy and angular 
momentum regimes affect this relationship, what the   
minimum $N$ is for which the standard mean free path approximations are 
still valid, etc.  

The second most noticable similarity between all Runs is 
an exponential decline in the collision probability
with increasing angular momentum that is steeper for larger $N$.  
This effect is small, however, relative to the total drop in 
the collision probability as $N$ decreases within a given Run (see 
Figures~\ref{fig:prob-merge-run1}, \ref{fig:prob-merge-run2}, 
and~\ref{fig:prob-merge-run3}).  
One possible explanation that is consistent with our results is 
that the number of close approaches per crossing time may decrease  
with increasing angular momentum more steeply for larger $N$ 
encounters.  

Our results are applicable to a regime of low angular momentum and high
absolute orbital energy.  
As the integration time increases with increasing angular momentum, we focussed
our attention on the low-angular momentum regime in this paper to
maximize the number of simulations performed for each Run, and thereby 
increase the statistical significance of our results.  

We cannot probe lower angular momentum encounters without 
violating our assumptions.  There are two reasons for this.  First, 
the lower boundary for the long-term stability of triple systems 
corresponds to a ratio 
between the inner and outer orbital separations $\gtrsim 7$
\citep{mardling01}.  Therefore, we cannot lower the total angular 
momentum by decreasing the semi-major axis of the outer orbit of the triple 
without simultaneously reducing the semi-major axis of its inner orbit.  
This brings us to our second requirement, namely that $q \ll a_0$ 
(where $a_0$ is the semi-major axis of the shortest-period orbit initially 
involved in the interaction 
and $q$ is the physical sizes of the objects involved in the encounter).  
We performed two additional Runs 
with $a_0 = 0.1$ AU to test the limit of the assumption $q \ll a_0$.  In 
both cases, the resulting chi-squared values found using the best-fits 
for our free parameters in Equation~\ref{eqn:prob-coll-N} 
were significantly larger than we found for our other Runs.  
We interpret this as being due to a breakdown of the requirement 
$q \ll a_0$.  Our results suggest that the assumption $q \ll a_0$ holds 
provided $a_0 \gtrsim 100q$.  This is only a rough estimate for the 
lower limit for the ratio $q/a_0$, and more work will be needed to better 
constrain its precise value.

In future work, we intend to address the $N$-dependence of the 
collision probability for higher angular momentum encounters, as well as 
different mass-ratios and eccentricities.  A non-circular orbit 
provides an additional free parameter that can be 
changed to affect the total angular momentum, but not the total 
energy.  Therefore, the addition of a non-zero eccentricty should in 
principle allow us to include 1+2 
encounters in our estimates for the collision probability using the 
same or similar normalization method (i.e. fixing the total angular 
momentum and energy) as adopted in this paper to compare between 
the different encounter types.  

\section{Summary} \label{summary}

In this paper, we perform a large suite of numerical scattering 
experiments to study the probability of a direct physical collision 
occurring during single-binary, binary-binary, single-triple, 
binary-triple, and triple-triple interactions.  We quantify
the dependence of the collision probability on the number of objects 
involved in the interaction $N$ for fixed total energy and angular 
momentum.  

Our results suggest that the collision probability 
increases approximately as $N^2$, for $N=4,5,6$.  
This result is consistent with the hypothesis that the
average number of close approaches per crossing time 
also scales as $N^2$, and we will fully investigate this possibility in a future study.
Interestingly, this same $N$-dependence is predicted by the mean free path 
approximation in the limit of very large $N$.  
This similarity is rather intriguing, and further work 
investigating the connection between these two regimes in $N$ will be highly valuable
for our understanding of collisional dynamics in the realm of not-so-small-$N$.

\appendix

\section{Schematic Representation of Liouville's
  Theorem} \label{appendix} 

In this appendix, we describe the construction of schematic 
diagrams that quantitatively describe the evolution of an 
interaction in energy-space.  

\subsection{Motivation} \label{motiv}

Previous numerical scattering experiments for 1+2 interactions have 
demonstrated that the probability for a dynamical encounter to result in 
a direct collision is approximately proportional to the number of close 
approaches per crossing time multiplied by the number of crossing times 
the system survives through \citep[e.g.][]{valtonen06}.  
This is the basis for the analytic model described in 
Section~\ref{collprob_Neq3} to predict the collision 
probability during resonant 1+2 encounters, and we have extended 
this formalism to also describe encounters with $N > 3$.
We briefly discuss the difficulties in determining the 
precise number of close approaches during an encounter
from an analysis of position-space in Section~\ref{discussion}.

Here we present an improved method for determining the number of close 
approaches per crossing time by analyzing encounters in energy-space.
There are three clear benefits to this technique, which we will expand upon 
below.  First, defining the condition for a close approach in terms of
energy rather than position maps each close approach to a nearly 
discrete change in energy, which is easily calculated.
Second, the energy-space diagrams presented in this paper 
contain quantitative information pertaining 
to the relative energies of the objects during each stage of the encounter,
whereas position-space diagrams do not (and are often very hard to interpret visually).  
Third, this method will potentially allow for a more general formalism
to describe the collision probability during a dynamical encounter.

Below, we outline our method and provide a few example scenarios.
We conclude by describing the future work that will be enabled by 
this energy-based approach.

\subsection{Liouville's Theorem} \label{louiville}

Liouville's Theorem is a key postulate in classical statistical
mechanics.  It states that the time evolution of the distribution
function corresponding to an Hamiltonian system is constant along any
trajectory in phase space \citep{liouville38}.  Said another way, the
density of points representing particles in \b{x},\b{p} phase space is
conserved as the system evolves in time, where \b{x} and \b{p} denote
the 3-D position and momentum vectors, respectively, of the
particles.  This is a remarkable and powerful result that can be
applied to any dynamically-evolving system provided the forces acting
on the particles are conservative and differentiable.  This last
condition excludes collisions due to the sudden introduction of
additional forces, such as tides, shocks, etc.  

As a system evolves dynamically, energy and angular momentum are 
exchanged between particles.  This diffusion in energy- and 
angular momentum-space governs the trajectories of the particles 
through position- and velocity-space.  Therefore, it is often 
the case that 
consideration of the evolution of a system in energy- and angular 
momentum-space, as opposed to position- and velocity-space, will 
more directly reveal the underlying physical processes that 
determine its outcome.  
As we will show in the subsequent sections, it 
follows that patterns in the dynamical evolution of an 
interaction are often most apparent in energy- and angular 
momentum-space.

\subsection{Diagrams for $N = 4$} \label{four}

We will begin by describing the rules for the construction of our 
schematic diagrams for the case $N = 4$.  
The total energy for a four-body encounter can be written:
\begin{equation}
\label{eqn:energy}
E = \sum_{i=1}^4 \frac{1}{2}m_iv_i^2 - \frac{1}{2}\sum_{i,j=1;i{\ne}j}^4 \frac{Gm_im_j}{r_{ij}},
\end{equation}
where $m_i$ is the mass of object $i$, $v_i$ is the speed of object
$i$ with respect to the centre of mass of the system, and $r_{ij}$ is
the distance separating objects $i$ and $j$.  We note that $r_{ij} = 
|\b{r}_i - \b{r}_j|$, where $\b{r}_i$ is the vector separating
object $i$ from the centre of mass of the system.  We re-write
Equation~\ref{eqn:energy} in the form:
\begin{equation}
\label{eqn:energy2}
E = \frac{1}{2} \sum_{i=1}^4 m_i \Big[ v_i^2 - \Big(
\sum_{j=1;j{\ne}i}^4 \frac{Gm_j}{r_{ij}} \Big)  \Big],
\end{equation}
or
\begin{equation}
\label{eqn:energy3}
E = \sum_{i=1}^4 E_i.
\end{equation}

Now, we can form a polygon by setting each of the angles equal
to:
\begin{equation}
\label{eqn:energy4}
\alpha_i = - \Big( \frac{E_i}{E} \Big) \times 360^{\circ}.
\end{equation}
(Note that the angle 360$^{\circ}$ here is a result of the more general formula
$180^{\circ} \times (N-2)$, when $N=4$.)  
As the system evolves, the total energy $E$ remains conserved but the 
individual energy terms $E_i$ will change.  This is manifested visually 
via the transformation of our 
schematic diagrams with successive time-steps.  The rules of our
diagrams are such that the individual angles of the polygon
change as the system evolves, while their sum remains 
constant.  

Bound objects are connected by solid lines, whereas 
objects that are unbound are connected by dashed lines.  Note that, 
if an object 
is unbound, then Equation~\ref{eqn:energy4} suggests that the angle
corresponding to its vertex will be negative.  
We will come back to this below.

\begin{figure}
\begin{center}
\includegraphics[width=\columnwidth]{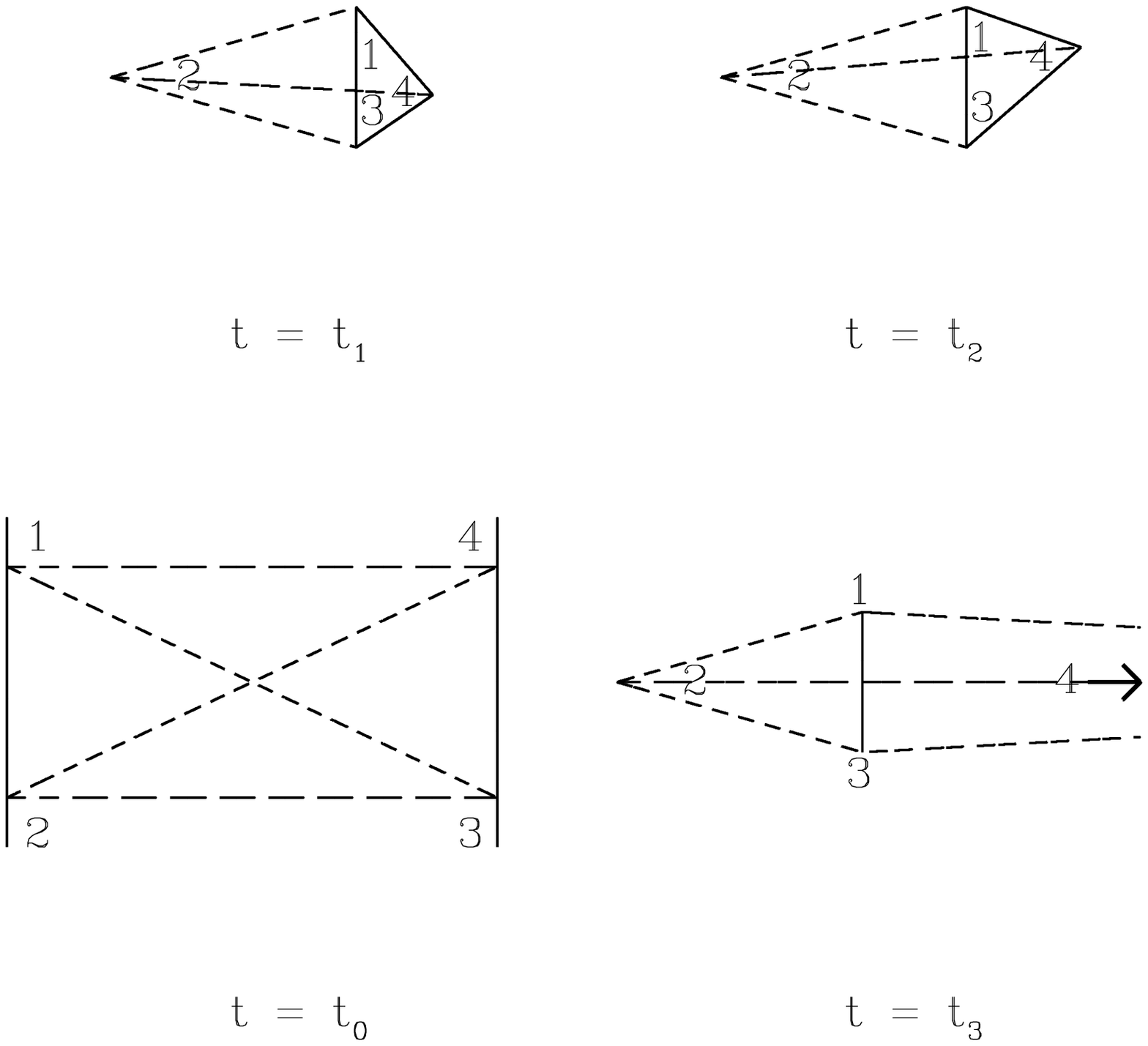}
\end{center}
\caption[Evolution in energy-space of a 2+2 interaction between 
identical binaries]{Schematic diagram 
showing the evolution in energy-space of a 2+2 interaction 
between identical binaries.  
The panels have been arranged in chronological order, 
starting at the lower left inset and rotating clockwise
(i.e. $t_0 < t_1 < t_2 < t_3$) .  A detailed
description of this interaction has been provided in the text.
The state of the system depicted in the inset corresponding to 
$t = t_2$ shows one example of the many transformations of 
our diagram that occur between ejection events.  This is 
due to the occurrence of several close approaches.
\label{fig:example-3}}
\end{figure}

\begin{figure}
\begin{center}
\includegraphics[width=\columnwidth]{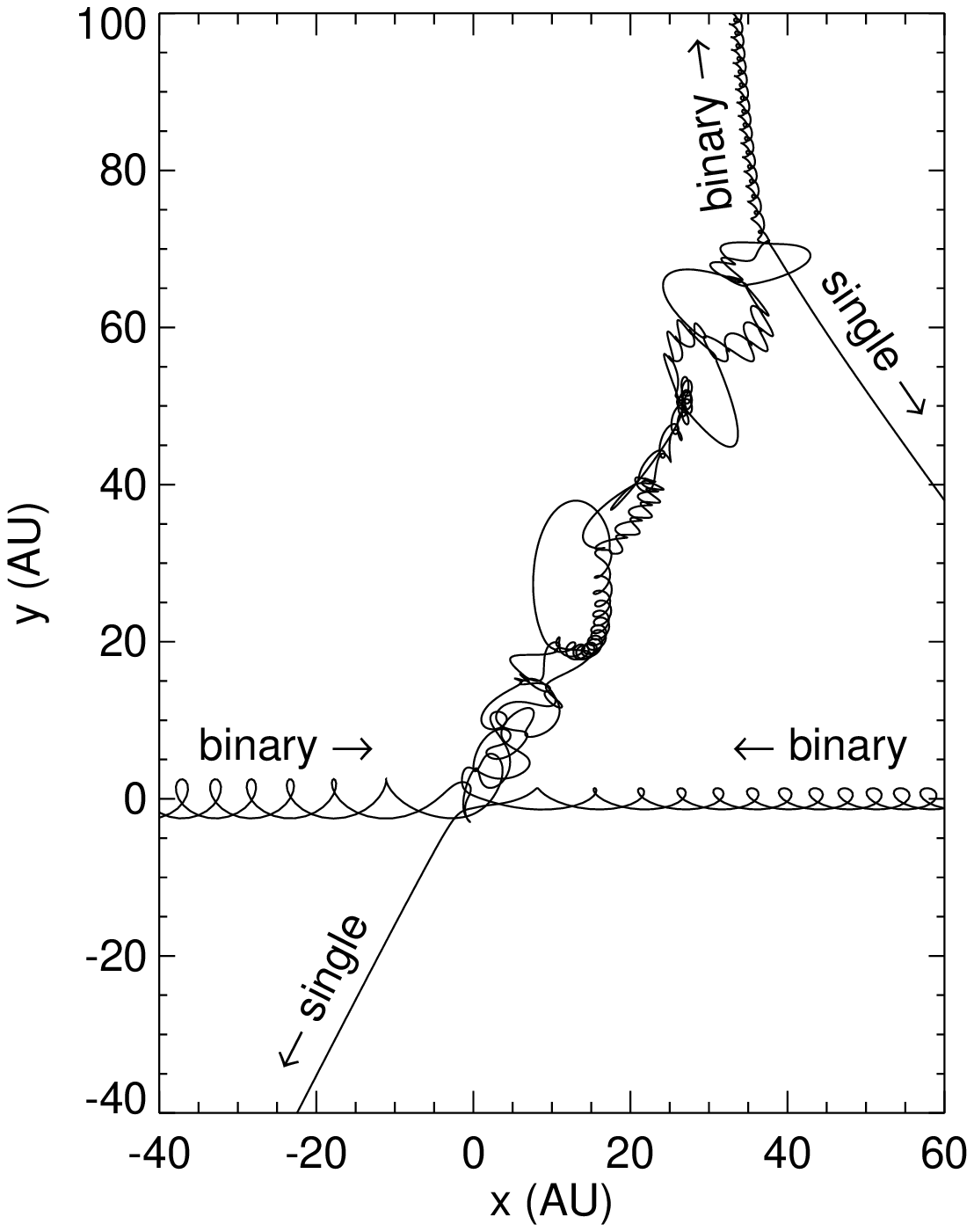}
\end{center}
\caption[Evolution of an 2+2 interaction in position-space projected onto 
the x-y plane]{Evolution of an 2+2 interaction in position-space projected onto
the x-y plane.  The binaries are composed of
1 M$_{\odot}$ stars, and have semi-major axes of 5 AU and
circular orbits.  The 
encounter involves an essentially parabolic collision between two identical 
binaries composed of equal mass stars, initially approaching from the 
left and right.  Distance is measured in units of the 
initial semi-major axes $a$ of the binaries.  When the encounter 
is over, two single stars and a binary emerge.
\label{fig:leonard1b}}
\end{figure}

To illustrate our methodology, we show the evolution
in energy-space of a 2+2 encounter in Figure~\ref{fig:example-3}.  
The corresponding position-space diagram is shown in 
Figure~\ref{fig:leonard1b}.  The latter figure shows the 
evolution of a 2+2 encounter in position-space projected onto the 
x-y plane.  In this interaction, two 
identical binaries that are composed of equal-mass (1 M$_{\odot}$) stars 
with semi-major axes of 5 AU approach from the left and right 
(we denote the initial time by $t = t_0$) to meet at the origin.  
In Figure~\ref{fig:example-3}, 
Stars 1 and 2 comprise one of the binaries, and Stars 3 and 4 comprise the other.  
Both of these bound pairs are connected with solid lines, respectively, 
while all remaining pairs of stars are initially unbound from each other, 
and therefore connected via dashed lines.  
Shortly after the binaries meet at the origin in Figure~\ref{fig:leonard1b}, 
one star (Star 2) is 
ejected from the system, and is expelled toward the lower left of the 
diagram ($t = t_1$).  Star 2 is now an unbound single object, and hence 
is connected to all other stars via dashed lines in the top left inset of 
Figure~\ref{fig:example-3}.  The angle corresponding 
to its vertex is also now negative, since Star 2 has a positive total energy.  

The three remaining stars then undergo a resonant 
interaction as they drift toward the upper right of the diagram in 
Figure~\ref{fig:leonard1b} due to 
conservation of momentum ($t = t_2$).  At roughly $(x,y) = (36,61)$, 
another star (Star 4) is ejected toward the lower right.  The left-over 
binary (composed of Stars 1 and 3) leaves the diagram at the upper right, 
with a semi-major 
axis (and hence orbital energy) that is smaller than those of the 
two initial binaries ($t = t_3$).  The final state of the system is depicted in 
energy-space in the lower right
inset of Figure~\ref{fig:example-3}.  Only Stars 1 and 3 are connected via a 
solid line, since they are the only two stars that remain bound.  The angles
corresponding to the vertices for Stars 2 and 4 are now negative, and
the angles corresponding to the vertices for Stars 1 and 3 are
both positive and obtuse (i.e. $> 180^{\circ}$).  
Note that, although described in detail here, the specific evolution of the encounter 
is quite difficult to follow visually in position-space without the help of 
the previous text.

\begin{figure}
\begin{center}
\includegraphics[width=\columnwidth]{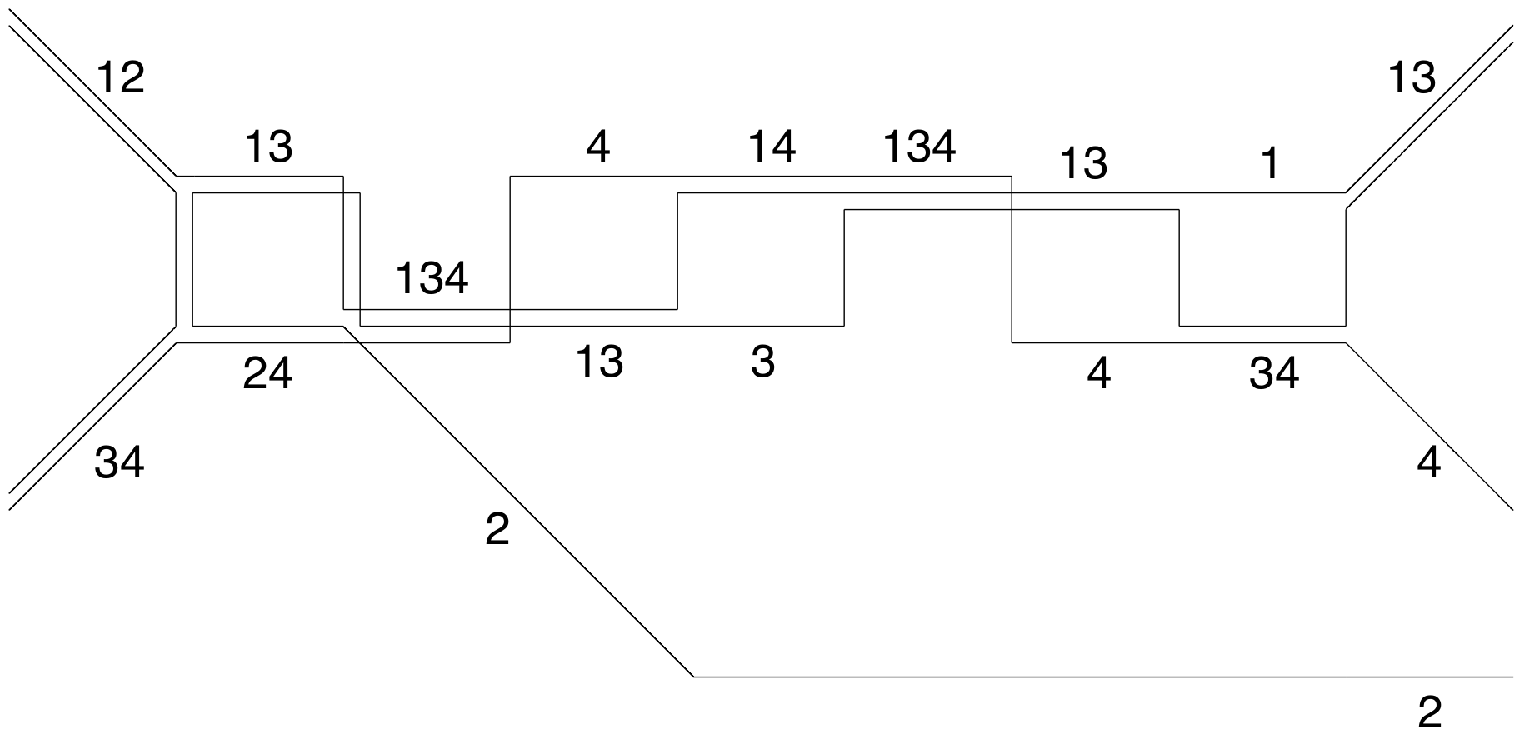}
\end{center}
\caption[Scattering diagram for the 2+2 interaction shown in
Figure~\ref{fig:leonard1b}]{Scattering diagram for the 
2+2 interaction shown in position-space in 
Figure~\ref{fig:leonard1b} and in energy-space in 
Figure~\ref{fig:example-3}.  The time evolution 
of the interaction proceeds from left to right.
\label{fig:scatdiag-3}}
\end{figure}

As a further step, we depict this encounter in Figure~\ref{fig:scatdiag-3} using 
the formalism for making scattering diagrams presented in \citet{hut83b}.  
In this plot, time increases from left to right, whereas vertical transitions 
denote the formation and/or destruction of temporary hierarchies in which 
two or more stars are in close proximity (i.e. strongly bound) to each other 
relative to all other stars.  
All four states of the system depicted in Figure~\ref{fig:example-3} are shown in 
Figure~\ref{fig:scatdiag-3}.  The initial (i.e. t $=$ t$_0$) and final (i.e. 
t $=$ t$_3$) states correspond to those shown at the far left and far 
right, respectively, of Figure~\ref{fig:scatdiag-3}.  The state shown at t $=$ t$_1$ 
corresponds to a time shortly after Star 2 has been ejected from the system.  
Note that, within the triangle formed from the vertices of Stars 1, 3, and 4
at t $=$ t$_1$, 
all of the angles are comparable.  This means that a temporary hierarchy has 
formed where Stars 1, 3, and 4 all have similar energies.  This is 
represented in Figure~\ref{fig:scatdiag-3} where the first 
convergence of all three lines (for Stars 1, 3, and 4) occurs simultaneously.  
This scenario occurs once 
more (i.e. at t $=$ t$_2$) before the interaction is eventually terminated by 
the escape of Star 4 from the system, leaving Stars 1 and 3 bound in a binary 
(which has a smaller orbital separation than the initial binaries going into 
the encounter).  

To summarize, after Star 2 is ejected, a three-body system remains that evolves 
via the formation 
of a temporary binary that mediates a series of successive ejections of the remaining 
single star (which can also be exchanged into and out of the temporary binary).  
Eventually, the single star receives a positive total energy and 
escapes from the system.

\subsection{Future Work} \label{future}

One primary application for this method is to determine the number of close 
approaches per crossing time for an encounter involving $N$ stars.  We outline 
this method here.  
A ``close approach'' can be defined using Equation~\ref{eqn:energy4} 
to convert a minimum distance (i.e. $q$ in Equation~\ref{eqn:prob-coll}) 
into a minimum angle.\footnote{Note that in the point-particle limit a 
``collision'' is considered to occur when two objects are within some 
minimum distance of each other.  Therefore, our energy-space diagrams can 
be used to depict collisions.}  
If we require that the distance of ``close approach'' is 
comparable to the physical sizes of the objects (which are assumed to be 
small relative to the characteristic size of the interaction region and 
the initial semi-major axes of any orbits going into the encounter), then 
these events correspond to times of near instantaneous and significant 
energy-exchange between two individual stars.  This is because the 
stars tend to be strongly accelerated or decelerated during very close 
approaches, so that significant energy and/or angular momentum can 
be exchanged.  It follows that a ``close approach'' will appear 
as a near discrete transformation of our energy-space diagrams, 
whereas the evolution appears more continuous and chaotic in 
position-space.  This is demonstrated in Figure 3 of 
\citet{hut83b} and also in Figure~\ref{fig:scatdiag-3}.  Therefore,
individual close approaches will correspond to high values for 
$d\alpha_i/dt$, where $\alpha_i$ is the angle corresponding to star $i$ in
an energy-space diagram.  Consequently, 
one can define a ``close approach'' as occurring when $d\alpha_i/dt > \eta$, 
where $\eta$ is a parameter that can be varied to specify the precise 
definition of a ``close approach''.
We intend to study the time-evolution of the quantity $d\alpha_i/dt$ in future 
work.

As is evident from Figure~\ref{fig:leonard1b}, 
the time evolution of the system is very difficult to observe in position-space, 
and counting the number of close approaches is essentially impossible from 
this type of a diagram.  In contrast, we can easily count ten (to first order) 
close approaches in Figure~\ref{fig:scatdiag-3}.  To at least first order, each 
time a close approach occurs, one of the the lines in a scattering diagram 
makes a vertical transition.  

This energy-space formalism is easily extended to depict the evolution 
of encounters involving more than 4 objects.  (In 
principle, there is no limit to the number of objects that 
can be described with this method.) 
Furthermore, we can, in principal, use this formalism to re-write the 
functional form for the collision 
probability shown in Equation~\ref{eqn:prob-coll-N} in terms of only 
the number of objects $N$, the total angular momentum, and energy.  This 
is because our method re-defines the condition for a ``close approach'' in 
terms of energy, instead of distance $q$.  The term $q/a_0$ in 
Equation~\ref{eqn:prob-coll-N} can thus be replaced with a new term that 
depends directly on energy.  This will further generalize our approach, and 
should be an improvement over the normalization used in this paper 
to compare between different encounter types within a given Run.

In addition, 
the same method we have presented to depict the evolution 
of the system in energy-space can be modified to depict its 
evolution in angular momentum-space.  The prescription for 
this is analogous to the formalism presented in the previous 
sections for energy-space.  In general, our technique can be 
used to depict the time evolution of any conserved quantity.

We intend to use our diagrams in future studies to simulate 
the dynamical evolution of small-$N$ gravitational interactions 
in both energy- and angular momentum-space.  This will be done in 
(accelerated) real time, so that we may simultaneously view the 
evolution of a system in 
position-, energy-, and angular momentum-space.  
This new tool will be a useful addition to studies with the 
over-arching goal of identifying trends in the evolution of 
dynamical interactions.  

\section*{Acknowledgments}

A very big thank you to John Fregeau for trouble-shooting assistance
when adapting the FEWBODY code, and for a critical read of our
manuscript.  Thank you to Alison Sills and Steinn Sigurdsson for
valuable feedback and guidance.  We would like to
thank the following people for useful discussions:  Hagai Perets,
Maureen van den Berg, Evert Glebbeek, and Bob Mathieu.
NL is funded by the European Space Agency (ESA) Postdoctoral Fellowship.
AMG is funded by the Lindheimer Fellowship at Northwestern University.

%\chapterbib

\bsp

\label{lastpage}


\begin{thebibliography}{99}

%\bibitem[\protect\citeauthoryear{Aarseth}{1973}]{aarseth73} Aarseth
%  S. J. 1973, Vistas in Astronomy, 15, 13
\bibitem[\protect\citeauthoryear{Anosova \& Orlov}{1986}]{anosova86} Anosova 
J. P., Orlov V. V. 1986, Soviet Astronomy, 30, 380
\bibitem[\protect\citeauthoryear{Bacon, Sigurdsson \& Davies}{1996}]{bacon96} 
Bacon D., Sigurdsson S., Davies M. B. 1996, MNRAS, 281, 830
%\bibitem[\protect\citeauthoryear{Brown, Geller \&
%    Kenyon}{2009}]{brown09} Brown W. R., Geller M. J., Kenyon
%  S. J. 2009, ApJ, 690, 1639
\bibitem[\protect\citeauthoryear{Davies}{1995}]{davies95} Davies M. B. 1995, 
MNRAS, 276, 887
\bibitem[\protect\citeauthoryear{Euler}{1772}]{euler1772} Euler L. 1772, 
Theoria Motuum Lunae, Typis Academiae Imperialis Scientiarum Petropoli 
(reprinted in Opera Omnia, Series 2, ed. L. Courvoisier, 22, 
Lausanne: Orell Fussli Turici, 1958)
\bibitem[\protect\citeauthoryear{Fregeau et al.}{2004}]{fregeau04}
  Fregeau J. M., Cheung P., Portegies Zwart S. F., Rasio F. A. 2004,
  MNRAS, 352, 1
%\bibitem[\protect\citeauthoryear{Harris}{1996, 2010 update}]{harris96}
%  Harris, W. E. 1996, AJ, 112, 1487 (2010 update)
\bibitem[\protect\citeauthoryear{Heggie}{1975}]{heggie75} Heggie
  D. C. 1975, MNRAS, 173, 729
\bibitem[\protect\citeauthoryear{Heggie \& Hut}{2003}]{heggie03}
  Heggie D. C., Hut P. 2003, The Gravitational Million-Body Problem:
  A Multidisciplinary Approach to Star Cluster Dynamics (Cambridge:
  Cambridge University Press)
\bibitem[\protect\citeauthoryear{Henon}{1969}]{henon69} Henon M. 1969, 
A\&A, 1, 223 
\bibitem[\protect\citeauthoryear{Hill}{1878}]{hill1878} Hill G. W. 1878, 
Reseaqrches in the lunar theory, Annals of the Journal of Mathematics, 1, 5 
%\bibitem[\protect\citeauthoryear{Hills}{1976}]{hills76} Hills J. 1976,
%  MNRAS, 175, 1p
\bibitem[\protect\citeauthoryear{Hut}{1983}]{hut83a} Hut P. 1983, ApJ, 268, 342
\bibitem[\protect\citeauthoryear{Hut \& Bahcall}{1983}]{hut83b} Hut P., Bahcall J. N. 
1983, ApJ, 268, 319
\bibitem[\protect\citeauthoryear{Hut}{1984}]{hut84} Hut P. 1984, ApJS, 55, 301
\bibitem[\protect\citeauthoryear{Ivanova et al.}{2006}]{ivanova06} Ivanova N.,
Heinke C. O., Rasio F. A., Taam R. E., Belczynski K., Fregeau J. M. 2006, 
MNRAS, 372, 1043
\bibitem[\protect\citeauthoryear{Ivanova et al.}{2008}]{ivanova08} Ivanova N., 
Heinke C. O., Rasio F. A., Belczynski K., Fregeau J. M. 2008, MNRAS, 386, 553
\bibitem[\protect\citeauthoryear{Jacobi}{1836}]{jacobi1836} Jacobi C. G. J. 1836, 
Sur le movement d'un point et sur un cas particulier du probleme des trois 
corps, Comptes Rendus, 3, 59
\bibitem[\protect\citeauthoryear{Lagrange}{1811}]{lagrange1811} Lagrange J. L. 1811, 
Mecanique Analytique, Paris
\bibitem[\protect\citeauthoryear{Leigh \& Sills}{2011}]{leigh11} Leigh N., 
Sills A. 2011, MNRAS, 410, 2370
\bibitem[\protect\citeauthoryear{Leonard}{1989}]{leonard89} Leonard
  P. J. T. 1989, AJ, 98, 217
%\bibitem[\protect\citeauthoryear{Leonard \& Linnell}{1992}]{leonard92}
%  Leonard P. J. T., Linnell A. P. 1992, AJ, 103, 1928
\bibitem[\protect\citeauthoryear{Liouville}{1838}]{liouville38}
  Liouville J. 1838, Journ. de Math., 3, 349
%\bibitem[\protect\citeauthoryear{Lombardi et al.}{2002}]{lombardi02}
%  Lombardi J. C., Warren J. S., Rasio F.A., Sills A., Warren
%  A. R. 2002, ApJ, 568, 939
\bibitem[\protect\citeauthoryear{Mardling}{2001}]{mardling01}
  Mardling R. A., 2001, in Podsiadlowski P., Rappaport S.,
  King A. R., D'Antona F., Burderi L., eds, ASP Conf. Ser. Vol. 229, Evolution of Binary
  and Multiple Star Systems. Astron. Soc. Pac., San Francisco, p. 101
\bibitem[\protect\citeauthoryear{Markwardt}{2008}]{markwardt08} Markwardt, C. B. 2008, 
``Non-Linear Least Squares Fitting in IDL with MPFIT,'' in proc. Astronomical Data Analysis 
Software and Systems XVIII, Quebec, Canada, ASP Conference Series, Vol. 411, eds. D. Bohlender, 
P. Dowler \& D. Durand (Astronomical Society of the Pacific: San Francisco), 
p. 251-254 (ISBN: 978-1-58381-702-5; http://adsabs.harvard.edu/abs/2009ASPC..411..251M )
\bibitem[\protect\citeauthoryear{McMillan}{1986}]{mcmillan86}
  McMillan S. L. W., 1986, ApJ, 306, 552
\bibitem[\protect\citeauthoryear{McMillan \& Hut}{1996}]{mcmillan96} McMillan S. L. W., 
Hut P. 1996, ApJ, 467, 348
\bibitem[\protect\citeauthoryear{Mikkola}{1983}]{mikkola83} Mikkola S. 1983, 
MNRAS, 203, 1107
\bibitem[\protect\citeauthoryear{Mikkola}{1984}]{mikkola84a} Mikkola S. 1984,
MNRAS, 207, 115
\bibitem[\protect\citeauthoryear{Mikkola}{1984}]{mikkola84b} Mikkola S. 1984,
MNRAS, 208, 75
\bibitem[\protect\citeauthoryear{Mikkola \& Valtonen}{1986}]{mikkola86} Mikkola S., 
Valtonen M. J. 1986, MNRAS, 223, 269
\bibitem[\protect\citeauthoryear{Mikkola \& Valtonen}{1990}]{mikkola90} Mikkola S.,
Valtonen M. J. 1990, ApJ, 348, 412
%\bibitem[\protect\citeauthoryear{Monaghan}{1976a}]{monaghan76a} Monaghan J. J. 
%1976, MNRAS, 176, 63
%\bibitem[\protect\citeauthoryear{Monaghan}{1976b}]{monaghan76b} MonaghanJ. J.
%1976, MNRAS, 177, 583
\bibitem[\protect\citeauthoryear{Mor\'{e}}{1978}]{more78} Mor\'{e}, J. 1978, ``The 
Levenberg-Marquardt Algorithm: Implementation and Theory,'' in Numerical Analysis, 
vol. 630, ed. G. A. Watson (Springer-Verlag: Berlin), p. 105 
(DOI: 10.1007/BFb0067690; http://www.springerlink.com/content/t6h81886n406/ )
\bibitem[\protect\citeauthoryear{Newton}{1686}]{newton1686} Newton,
  I. 1760, Philosophiae Naturalis Principia Mathematica (Trinity
  College:  Regalis Societatis Praesses)
%\bibitem[\protect\citeauthoryear{Paczynski}{1967}]{paczynski67}
%  Paczynski B. 1967, Acta Astr., 17, 287
%\bibitem[\protect\citeauthoryear{Perets \& Fabrycky}{2009}]{perets09}
%  Perets H. B., Fabrycky D. C. 2009, ApJ, 697, 1048
\bibitem[\protect\citeauthoryear{Poincare}{1892}]{poincare1892} Poincare H. 
1892, Les methodes nouvelles de la mechanique celeste (Paris: Gauthier-Villars) 
%\bibitem[\protect\citeauthoryear{Sandage}{1953}]{sandage53}
%  Sandage A. R. 1953, AJ, 58, 61
%\bibitem[\protect\citeauthoryear{Sandquist et al.}{2003}]{sandquist03}
%  Sandquist E. L., Latham D. W., Shetrone M. D., Milone
%  A. A. E., 2003, AJ, 125, 810
\bibitem[\protect\citeauthoryear{Saslaw}{1974}]{saslaw74} Saslaw W. C., 
Valtonen M. J., Aarseth S. J. 1974, ApJ, 190, 253
\bibitem[\protect\citeauthoryear{Sigurdsson \&
    Phinney}{1993}]{sigurdsson93} Sigurdsson S., Phinney E. S. 1993,
  ApJ, 415, 631
\bibitem[\protect\citeauthoryear{Sigurdsson \&
    Phinney}{1995}]{sigurdsson95} Sigurdsson S., Phinney E. S. 1995,
  ApJS, 99, 609
%\bibitem[\protect\citeauthoryear{Sills et al.}{2001}]{sills01} Sills
%  A. R., Faber J. A., Lombardi J. C., Rasio F. A., Waren A. R. 2001,
%  ApJ, 548, 323
\bibitem[\protect\citeauthoryear{Spitzer}{1987}]{spitzer87} Spitzer
  L. 1987, Dynamical Evolution of Globular Clusters (Princeton:
  Princeton University Press)
\bibitem[\protect\citeauthoryear{Sundman}{1912}]{sundman1912} Sundman K. F. 
1912, Acta Math., 36, 105
%``Mémoire sur le problème des trois corps.'' Acta Math. 36, 105 
\bibitem[\protect\citeauthoryear{Szebehely}{1967}]{szebehely67} Szebehely V. 
1967, The Theory of Orbits, New York:  Academic Press
\bibitem[\protect\citeauthoryear{Valtonen}{1974}]{valtonen74} Valtonen M. J. 
1974, Statistics of three-body experiments, in Stability of the Solar System 
and of Small Stellar Systems, Proc. IAU Symp. 62, ed. Y. Kozai, Dordrecht: 
Reidel, 211
\bibitem[\protect\citeauthoryear{Valtonen et al.}{1994}]{valtonen94} 
Valtonen M. J., Mikkola S., Heinamaki P., Valtonen H. 1994, ApJS, 95, 69
\bibitem[\protect\citeauthoryear{Valtonen \&
    Karttunen}{2006}]{valtonen06} Valtonen M., Karttunen H. 2006, The
  Three-Body Problem (Cambridge: Cambridge University Press)
%\bibitem[\protect\citeauthoryear{van den Berg et
%    al.}{2001}]{vandenberg01} van den Berg M., Orosz J., Verbunt
%  F., Stassun K., 2001, A\&A, 375, 375
%\bibitem[\protect\citeauthoryear{Verbunt \& Lewin}{2005}]{verbunt05}
%  Verbunt F., Lewin W. H. G. Globular Cluster X-Ray Sources, In:
%  Compact Stellar X-ray Sources, eds. W. H. G. Lewin and M. van der
%  Klis (Cambridge:  Cambridge Univeresity Press), 2006, pp. 341-379 

\end{thebibliography}
\end{document}